\documentstyle[twoside,fleqn,espcrc2]{article}
\newcommand{\fee}{\varphi}

\newcommand{\AmS}{{\protect\the\textfont2
  A\kern-.1667em\lower.5ex\hbox{M}\kern-.125emS}}

\title{Composite Weak Bosons: a Lattice Analysis}

\author{A. Galli\\[0.1cm]
{\em Paul Scherrer Institute, CH-5231 Villigen PSI, Switzerland}}

\begin{document}

\begin{abstract}
We present a lattice analysis of a
confining Yang-Mills theory without Goldstone boson.
We have analytically investigated the
model by a strong coupling expansion and
by an intensive lattice Monte Carlo simulation using standard lattice QCD
methods.
We show that this
theory is an interesting candidate for describing weak bosons as composite
particles.
\end{abstract}

\maketitle
The Standard Model \cite{SM} (SM) describes the strong, weak and
electromagnetic
interactions by a gauge theory based on the group
$G_{SM}=SU(3)\times SU(2)\times U(1)$
which is broken by the Higgs mechanism to $SU(3)\times U(1)$. The theory is
essentially determined once the matter fields and their transformation
under the local gauge transformations of $G_{SM}$ are specified. The matter
fields (leptons and quarks) and the Higgs boson are considered to be
elementary. They interact with each other by the exchange of gauge bosons which
are also considered to be elementary. The structure of the SM has been
phenomenologically confirmed to high accuracy.\\
In spite of the beautiful corroboration of the SM by experiments a natural
questions arises: How elementary are the leptons, the quarks, the Higgs bosons
and the gauge bosons?
The idea that the SM itself is an effective theory of another, more
fundamental, theory,
where quarks, leptons and bosons are composites of more fundamental
fields, is almost as old as the SM itself. The idea of quark and lepton
compositeness is motivated by the observed connection between quarks and
leptons, by the generation puzzle and by the existence of too many parameters
in the SM. The Higgs compositeness is motivated by the fine tuning problem.
The W and Z compositeness is motivated by their relation to a composite Higgs
and by the observation that all short-range interactions are residual
interactions of a more fundamental long-range interaction.\\
The substructures, the new fundamental fields, are supposed to carry a new
internal quantum number (which we refer to {\em hypercolor}) and the quarks,
leptons and bosons are hypercolorless composite systems of them. The binding of
the
substructures due to hypercolor is viewed as an analogy with the color
confinement
mechanism of QCD. However,
since the SM spectrum is different from the hadron spectrum, the hypercolor
interaction has to be described by a strongly coupled Yang-Mills theory
{\em different} from QCD.\\
Several models treat the quarks, leptons and bosons as composite systems.
Today a conspicuous number of theorems
exist which have ruled out most of the existing models and radically restricted
the possibilities of constructing realistic composite models
\cite{1}. One particular model has survived:
The Yang-Mills theory without
Goldstone bosons \cite{11}.
This model is a usual confining Yang-Mills theory with $SU(2)$ local
hypercolor gauge group, $SU(2)$ global isospin group and generalized Majorana
fermions in the fundamental representation of the local and global symmetry
groups. \\

We consider a gauge theory whose fermion content is represented by a Weyl
spinor
$F_{\alpha,a}^A(x)$. Here $\alpha$ denotes the (undotted) spinor
index ($\alpha=1,2$),
$A$
denotes the fundamental representation index of a global SU(2) isospin group
($A=1,2$) and $a$ denotes the fundamental representation index of the local
SU(2) hypercolor gauge group ($a=1,2$). We introduce the generalized
Majorana spinor $\psi$ starting from the Weyl spinors $F$ and its
conjugate $F^\dagger$
\begin{equation}
\psi(x)=\left(\begin{array}{c}F(x)\\QF^\dagger(x)\end{array}\right)=
\left(\begin{array}{cc}1&0\\0&Q\end{array}\right)\fee(x)
\end{equation}
and its adjoint
\begin{equation}
\bar\psi(x)=(F^T(x)Q,F^\dagger(x))=\fee^T(x)
\left(\begin{array}{cc}Q&0\\0&1\end{array}\right).
\end{equation}
The matrix $Q$ represents the antisymmetric matrix in spin, hypercolor and
isospin space. Of course the fields $\psi$ and $\bar{\psi}$ are
not independent fields.
The choice of the global isospin group $SU(2)$ and of the local hypercolor
group
$SU(2)$ allows us to write a gauge invariant mass term
for the generalized Majorana
fermion fields
\begin{equation}
\bar{\psi}\psi=FQF+F^\dagger QF^\dagger.
\end{equation}
Note that this choice is unique if one deals with Majorana fermions.
Because of the existence of the mass term we can define the Yang-Mills action
on the lattice in Euclidean space in the form of a Wilson action.
For further details we refer to ref. \cite{13}.\\

To be precise, this model considers the photon to remain elementary and
switched off. The weak gauge bosons $W^{\pm}$ and $Z^0$ then form a mass
degenerate triplet. The vector isotriplet bound state of the substructure
represents the W-boson triplet.
To be viable  a composite model of the
weak bosons has to reproduce the known weak boson spectrum: the lightest
bound states have to be the W-bosons and heavier bound states have to lie
in an experimentally unexplored energy range.
The only possibility of having a Yang-Mills theory which reproduces the weak
boson spectrum is to choose the degrees of freedom in a way
that they naturally avoid bound states lighter than the vector isotriplet
of the theory which characterizes the W-boson triplet.
This is possible if
the unwanted light bound states which naturally show up as
Goldstone bosons or pseudo Goldstone bosons in many models (like, for example,
a pseudoscalar isomultiplet, which would be the pion analogue of QCD) are
avoided. The choice of the generalized Majorana
fermions in this model avoids the $SU_A(2)$ global chiral symmetry of the
Yang-Mills Lagrangian because left- and right-handed degrees of freedom are not
independent. The axial current (which would generate the $SU_A(2)$ chiral
symmetry) does not exist and it is not possible to have a
breaking of $SU_A(2)$ with the related low lying Goldstone bosons.
In fact, the pseudoscalar isotriplet vanishes by the Pauli principle
(it is a symmetric combination of Grassmann variables).\\

Because of the confining character of this theory, we need
non-perturbative methods to make predictions. It is important that the
fermion theory under discussion can be defined by a gauge invariant lattice
regularization. A lattice regularization \`a la Wilson \cite{12}
is possible because the choice of the
isospin group $SU(2)$ allows us to replace the Dirac mass term and the
Dirac-type Wilson term by a hypercolor gauge invariant Majorana type
expression.\\
An extensive strong coupling expansion analysis \cite{13} of the
spectrum of this theory has shown that the spin one isotriplet bound state
(the right quantum number to represent the W-boson of the SM) could be the
lightest state if the pseudoscalar isosinglet acquires a mass by the
chiral anomaly in analogy with the $\eta'$ in QCD. \\
We also
calculated the spectrum of the lightest bound states by a quenched
Monte Carlo simulation \cite{MC} and we showed that the vector
isotriplet bound state of this theory is the lightest one.
We have performed our Monte Carlo simulations using the
following standard technique of lattice QCD:
\begin{enumerate}
\item We used the quenched approximation. As in QCD we assume that the
quenched approximation is reasonable also in our model.
\item To generate the quenched gauge configurations we used the heat-bath and
over relaxed updating (1 heat-bath sweep for 6 over relaxed sweeps).
\item We used the symmetrized Peskin's formula \cite{peskin} and the cooling
algorithm \cite{cool} to
measure the topological charge of the gauge configurations. Topological
non-trivial configurations are needed to evaluate the chiral anomaly
contribution to the mass of the pseudoscalar isosinglet bound state.
\item To invert the fermion matrix we used the minimal residual
and the conjugate gradient algorithms.
\item We used the smearing technique \cite{smearing}
(PSI-Wuppertal smearing) to obtain early plateaux in the local
masses.
\end{enumerate}
To connect the measured quantities to physics we have to fix the lattice
spacing $a$.
For setting the lattice spacing $a$ we identify the experimental value of
the $W-$boson triplet $M_W=80$ GeV with the value of the mass of the
vector isotriplet determined from the simulations and extrapolated to the
chiral limit.
We expect that the mass of the substructure is much smaller than a
typical binding energy, therefore in analogy with QCD we
extrapolated the bound state masses to the chiral limit, determined by the
critical hopping parameter.
In our model there is {\em no} Goldstone boson which can identify the chiral
limit. However,
the critical hopping parameter can be evaluated by assuming that the
pseudoscalar isosinglet behaves like a Goldstone boson when the contribution of
the chiral anomaly is switched off and therefore it is massless. \\

We have performed
the simulations on different lattices and with
different values of $\beta$ to control
finite $a$ and volume effects.
In some simulations we have calculated the masses of bound states
without the
contribution of the chiral anomaly. These simulations confirmed the results of
the strong coupling expansion.
In other simulations we have
used and improved the method of ref. \cite{ito}
to evaluate the chiral anomaly
contribution to the mass of the pseudoscalar isosinglet bound state. This
computation required the evaluation of disconnected fermion loops and the
generation of topological non-trivial gauge configurations. The mass of the
pseudoscalar isosinglet turned out to be larger than the vector isotriplet
mass.\\

As a main result our lattice simulations has shown that the vector isotriplet
bound state, which represents the weak boson triplet, is the lightest bound
state in our model as it should be for any viable candidate of electroweak
composite model. In addition we have also predicted the masses of the first
bound states which are
heavier than the vector isotriplet representing the weak boson
triplet. These bound states are an additional vector isotriplet, a vector
isosinglet and a pseudoscalar isosinglet with masses in the range of a few
hundred GeV. These predictions open new interesting experimental perspectives
at
LEPII and LHC.

\end{document}